\documentclass[
    aps,
    pra,
    amsmath,
    amssymb,
    superscriptaddress,
    showpacs,
    floatfix,
    intlimits,
    reprint
]{revtex4-1}

\usepackage[english]{babel}
\usepackage[utf8]{inputenc}

\usepackage{amsthm}
\usepackage{mathtools}
\usepackage{physics}
\usepackage{xcolor}
\usepackage{graphicx}
\usepackage[left=23mm,right=13mm,top=35mm,columnsep=15pt]{geometry} 
\usepackage{adjustbox}
\usepackage{placeins}
\usepackage[T1]{fontenc}
\usepackage{lipsum}
\usepackage{csquotes}
\usepackage{orcidlink}

\begin{document}

\title{Quantum differential interference contrast microscope}

\author{Daniel F. Urrego~\orcidlink{0000-0002-9067-7909}}
\email{daniel.urrego@icfo.eu}
\affiliation{ICFO – Institut de Ciències Fotòniques, The Barcelona Institute of Science and Technology, 08860 Castelldefels Spain}

\author{M. H. M. Passos~\orcidlink{0000-0002-1571-5338}}
\altaffiliation{Currently at: Centro Brasileiro de Pesquisas Físicas, Rua Dr. Xavier Sigaud, 150, Urca, Rio de Janeiro, 22290-180, RJ, Brazil.}
\affiliation{ICFO – Institut de Ciències Fotòniques, The Barcelona Institute of Science and Technology, 08860 Castelldefels Spain}

\author{Francisco Bernal-Texca~\orcidlink{0000-0002-6104-6913}} 
\affiliation{ICFO – Institut de Ciències Fotòniques, The Barcelona Institute of Science and Technology, 08860 Castelldefels Spain}

\author{Juan P. Torres~\orcidlink{0000-0002-4454-6676}}
\email{juanp.torres@icfo.eu}
\affiliation{ICFO – Institut de Ciències Fotòniques, The Barcelona Institute of Science and Technology, 08860 Castelldefels}%
\affiliation{Department of Signal Theory and Communications, Universitat Politecnica de Catalunya, 08034 Barcelona, Spain}%

\begin{abstract}
\textbf{Optical sensing and imaging of transparent unstained biological and material specimen makes use of optical interference to translate phase shifts into intensity measurements, that are readily available in the optical domain. This is the case of Differential Interference Contrast (DIC) microscopes. However, environmental disturbances such as acoustic vibrations, air currents, temperature shifts, and mechanical drifts cause random phase noise that degrades the precision attainable in the estimation of the unknown phase shift. Although there are tools to compensate for these random phase drifts, such using common path interference techniques or phase stabilization systems, these are not always an option and its introduction is generally at the expense of device simplicity and robustness. Here we introduce a quantum version of Differential Interference Contrast microscopy (qDIC), that makes use of quantum interference for polarization analysis. When compared with standard schemes that make use of coherent light, our approach shows that the transfer of classical DIC microscopy into the quantum domain offers a route towards quantum-enhanced resilience to environmental disturbances, allowing the enhancement of phase sensitivity without sacrificing the simplicity and robustness of the optical setup.}
\end{abstract}

\flushbottom
\maketitle
\thispagestyle{empty}
%\linenumbers

%\section{Introduction}
\label{intro}
Differential Interference Contrast (DIC) microscopy, also known as Nomarski microscopy, is a label-free optical imaging technique that visualizes optical thickness variations of a phase object by transforming local thickness gradients into optical path differences~\cite{murphy2003book}. It has a large range of applications, from measuring nanometric-thick profiles of dielectric nano-particles \cite{hamilton2022} to measuring cell lines \cite{yasuhiko2021} or the size of single and double protein layers in microarray plates covered with biomarker targets \cite{Terborg2016}.

In DIC, the optical beam that illuminates the object under study is diagonally polarized. With the help of a Savart plate or a Wollaston prism, the specimen is sampled by a pair of closely spaced optical beams with orthogonal polarizations. The two beams acquire different phase shifts since they traverse the sample at different, but close, locations. As a result, after recombination of the two beams with orthogonal polarizations into a single beam, the state of polarization of the output optical beam changes with respect to the input state of polarization. 

\begin{figure*}[t!]
 \centering
 \includegraphics[width=0.9\linewidth]{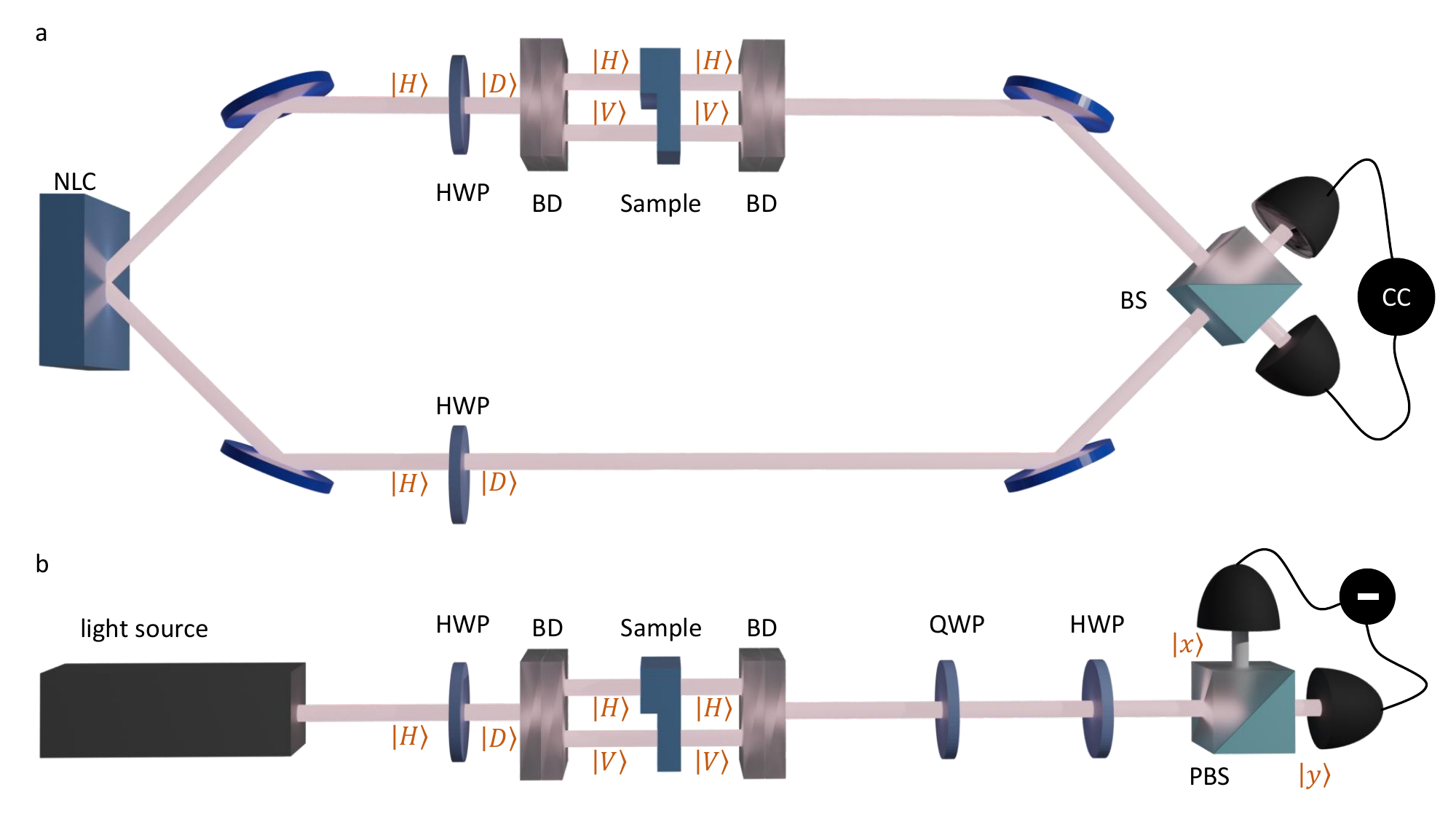}
 \caption{(a) Quantum and (b) classical experimental schemes of a Differential Interference Contrast (DIC) Microscopy. $|H \rangle$, $|V \rangle$, and $|D \rangle$ designate states of polarization: horizontal, vertical and diagonal. $|x \rangle$ and $|y\rangle$ designate polarization states onto which the output signal in classical DIC is projected before measurement. SP: Savart plate; QWP: quarter-wave plate; HWP: Half-wave plate; PBS: Polarizing beam splitter; BS: Beam splitter; SPCM: single-photon counting module; cc: coincidence counts; NLC: nonlinear crystal. The symbol $-$ represents subtraction of the signals detected in the two output ports of the PBS. }
\label{experimental_setup1}
\end{figure*}

The standard experimental setup used to detect polarization changes in DIC, and thus the corresponding phase shifts, it makes use of a half-wave plate (HWP) and a polarizing beam splitter (PBS) to project the output optical beam onto a certain polarization state \cite{Terborg2016}. The aim is to detect increasingly smaller optical path differences. Ultra-small phase shifts, of up to $\sim 10$ nanoradians have been measured \cite{fejer2006,matsuda2009} using common path interference techniques, which keep the phase stability of the interferometer by making the interfering optical beams traverse the same optical paths.

However, the use of common path interferometry schemes can be cumbersome, since they require specialized optical elements, which may make the system more bulky and harder to align. For certain applications it might not even be an option, since there are inherently more than one optical path, as it is the case of optical coherence tomography. Although DIC is close to a common path interferometer, since the interfering beams share a large part of the optical path, still it intrinsically contains two different paths for photons with orthogonal polarizations. The use of phase control systems to compensate for random phase drifts adds complexity to optical setups, especially when aiming at measuring the tiny phase changes involved in imaging nanostructures.

How the presence of unavoidable random phase fluctuations affect the precision of phase estimation, and thus the accuracy and robustness of DIC? Here we will show that the use of quantum interference for phase estimation in a DIC imaging scheme, what we can name as a quantum Differential Interference Contrast (qDIC) microscope, is more resilient to the presence of random phase fluctuations than a standard scheme that makes use of coherent state illumination, what we might call illumination with {\it classical} light. 

Quantum sensing and imaging has emerged as a promising field to enhance imaging and sensing capabilities \cite{GilaberteBasset2019}, for instance, providing strategies that offer optimum precision of phase estimation~\cite{pezze2017}, or more precision per photon in scenarios that requires very weakly invasive techniques \cite{Perarnau2021}. Several schemes have been implemented using intrinsically quantum properties such as photonic entanglement \cite{Ono2013}.

Imaging systems based on the use of quantum interference in a Hong-Ou-Mandel configuration~\cite{Hong1987} have been implemented~\cite{Chrapkiewicz2016,Lyons2018,Ndagano2022}. However, it is not evident that these techniques provide a {\it practical}, even a {\it fundamental}, quantum advantage in terms of enhanced sensitivity when compared with classical counterpart schemes. For instance, for measuring in a reflection geometry the size of a cliff-like nanostructure (optical trench) of height $h \sim 60$ nm, the attosecond-resolution Hong-Ou-Mandel interferometer demonstrated in~\cite{Lyons2018} requires to control/measure delays of $\sim 2 h /c=200~ as$. Apart from the difficulty of controlling/measuring such tiny temporal delays, increased precision requires the use of photons with a large bandwidth, which make it more susceptible to random phase noise.

In order to claim that a quantum technology provides a quantum advantage, it needs to make use of an inherently quantum feature that has no equivalence in the classical world. In our experiments with a {\it quantum} Differential Interference Constrast microscope, we make use of quantum coherence, the interference of two events that can happen, leading to the same experimental outcome. However, interference takes place only if there is no way to distinguish what event is indeed happening. This is in contrast to classical coherence, the basic idea behind a {\it classical} Differential Interference Contrast microscope, where interference takes place between streams of photons with orthogonal polarizations that acquire different phase shifts.
 
Most times is not straightforward how to make a fair comparison between quantum and classical systems, an analysis that could identify where it resides the quantum advantage in an imaging system, even to show that there is indeed any advantage at all~\cite{tang2021,tang2022}. For instance, some features of the so-called {\it quantum optical coherence tomography}~\cite{Nasr2003}, which were thought to require the use of frequency entangled paired photons, can be mimicked using light with other types of non-entangled correlations~\cite{shapiro2006,shapiro2010}. 

Here we put forward an imaging technique that manifestly provides a quantum-enhanced sensitivity advantage. We compare our approach with a classical system that has been used to measure phase shifts of up to $\sim 10$~nm in another context. Even though the classical system can provide better sensitivity in an {\it ideal} scenario with no environmental disturbances, quantum DIC turns up to be more resilience to the presence of random phase noise that its counterpart classical DIC. The use of quantum interference between single photons provides better sensitivity than classical coherence between optical beams under the presence of loss and random phase noise. 

\section*{Sensitivity analysis of the quantum and classical DIC}
Figure~\ref{experimental_setup1}(a) shows the experimental scheme of a {\it quantum} Differential Interference Contrast microscope. The quantum photonic state after interaction with the sample is a two-photon state that in the single-mode approximation can be written as
\begin{eqnarray}
& & \ket{\Psi_2}=\frac{1}{2}\,\Big[ a_H^{\dagger} \exp(i \varphi_H)+ a_V^{\dagger} \exp(i \varphi_V) \Big] \nonumber \\
& & \times \Big[ b_H^{\dagger} + b_V^{\dagger} \exp(i \theta) \Big] \ket{\text{vac}} 
\label{Psi_2}
\end{eqnarray}
where $a_{H,V}$ designate creation operators for photons illuminating the sample, $b_{H,V}$ designate creation operators for the reference photons, $\varphi_{H,V}$ are the phase shifts acquired by the orthogonal polarizations and $\theta$ is a reference phase. The setup shown in Figure~\ref{experimental_setup1}(a) reveals the degree of quantum coherence through the Hong-Ou-Mandel (HOM) effect~\cite{Hong1987}, a genuinely quantum phenomenon with no classical counterpart. The two events that interfere are both photons, each entering the beam splitter through a different input port, being reflected or both photons being transmitted. Both possibilities leads to the same experimental output, the detection of a photon coincidence count between photons in the two output ports of the beam splitter. 

For the sake of comparison, we consider the classical counterpart system. Figure~\ref{experimental_setup1}(b) shows the standard setup of a Differential Interference Contrast (DIC) microscope. It makes use of coherent light, so the quantum state of light after interaction with the sample is 
\begin{equation}
 \ket{\Psi_1}=\Bigg|\frac{\alpha \exp \left( i \varphi_H \right)}{\sqrt{2}}\Bigg\rangle_H\,\Bigg|\frac{\alpha \exp \left( i \varphi_V \right)}{\sqrt{2}}\Bigg\rangle_V
 \label{Psi_1}
\end{equation}
where $N_0=|\alpha|^2$ is the number of photons that illuminate the sample. Since this DIC microscope makes use of coherent light, it can be loosely named as {\it classical} DIC. The signal that yields the spatial map of optical path differences generated by a sample is the difference signal $S$ obtained from signals $N_1$ and $N_2$ detected in both output ports of the polarizing beam splitter, i.e., $S=N_1-N_2$. This is the setup that some optical schemes has used to measure in other scenarios some ultra-small phase shifts~\cite{fejer2006,matsuda2009}.

As test sample we consider a cliff-like structure as shown in Figures~\ref{experimental_setup1}(a) and (b). For a structure with step height $h$ and made of a material with refractive index $n_0$, the phase shift between orthogonal polarizations is $\varphi_0=\varphi_H-\varphi_V=k_0 (n_0-1)d$, where $k_0=2\pi/\lambda$ is the wavenumber and $\lambda$ is the wavelength. 

We first consider the quantum Cramer-Rao bound (qCRB), the bound to the best precision $\sigma_{\varphi}$ in the estimation of the unknown phase $\varphi$ that can be attained given the quantum state, for any possible experiment one can devise. In Supplementary Section I we calculate the qCRM. We find that for classical DIC, $\sigma_{\varphi} \geq 1/\sqrt{N_0}$ and for quantum DIC $\sigma_{\varphi} \geq 1/\sqrt{M}$, where $M$ is the number of trials of the experiment. Since in DIC $N_0$ photons interacts with the sample on average, and in qDIC $M$ photons interact with the sample, both DIC schemes share a similar quantum Cramer-Rao bound. 

In order make an overall comparison between the sensitivity of quantum DIC and classical DIC, we take into account differential losses between the horizontal and vertical polarization channels (transmissivities $T_1$ and $T_2$), as well as the efficiency of photo-detection ($T_d$). One important aspect of the performance of any interferometer is the resilience to random phase noise, caused by environmental disturbances such as acoustic vibrations, air currents, temperature shifts, and mechanical drifts, that degrades the phase stability of the interferometer. To take into account this, we consider that there is path imbalance $x$ in the interferometer between the horizontal and vertical polarization channels, that changes randomly with probability distribution
\begin{equation}
P(x)=\frac{1}{\left( 2\pi W_0^2 \right)^{1/2}} \exp \Big[ -\frac{x^2}{2W_0^2}\Big]
\end{equation}
where $W_0$ is the effective width of the random fluctuations. 

\begin{figure}[t!]
 \centering
 \includegraphics[width=\linewidth]{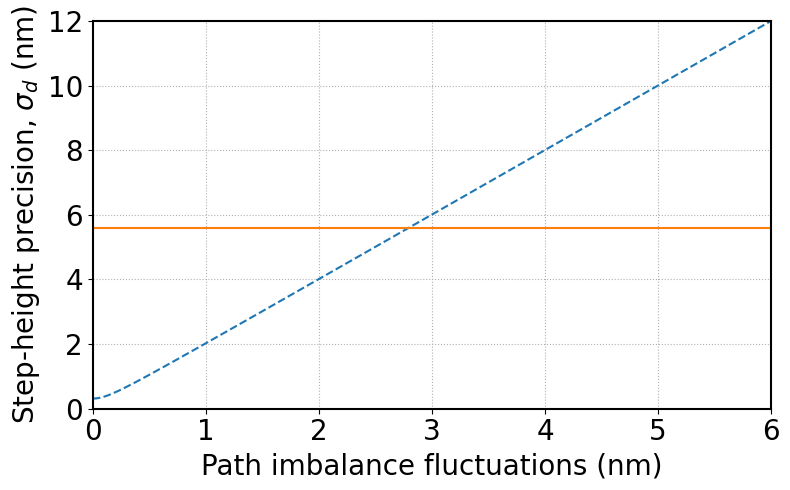}
 \caption{\textbf{Comparison of the sensitivity (typical deviation $\sigma_h$) between classical and quantum DIC}. Solid orange line: sensitivity of quantum DIC; Blue dashed line: sensitivity of classical DIC. As test sample, we consider a cliff-like structure (optical trench) where we want to estimate a step height of $h=40$ nm. Average number of photons of the coherent state: $N_0=10^6$; Number of paired photons we measure: $M=10^5$; Bandwidth of SPDC: $\Delta \lambda_s=3$ nm. We make use of Eqs. (\ref{sigma_qDIC}) and (\ref{sigma_cDIC}).}
\label{qadvantage}
\end{figure}

In Supplementary Section II we calculate the precision of phase estimation for quantum DIC based on the setup shown in Figure~\ref{experimental_setup1}(a). We obtain that the precision is
\begin{eqnarray}
 & & \sigma_{\varphi}^2= \frac{1+4 \sigma^2 \Delta^2}{T_1 T_2 T_d^2 M\, \alpha^2\,\sin^2 \varphi_0} \Big[ 1-\frac{\alpha}{\sqrt{1+4 \sigma^2 \Delta^2}} \cos \varphi_0 \Big] \nonumber \\
 & & \times \Big[ 4-T_1 T_2 T_d^2+\frac{\alpha\, T_1 T_2 T_d^2}{\sqrt{1+4 \sigma^2 \Delta^2}} \cos \varphi_0\Big] 
 \label{sigma_qDIC}
\end{eqnarray}
By tuning the relative arrival time (delay) between the photons entering the beam splitter, one observe the so-called HOM dip, in which the coincidence count rate reaches a minimum, that ideally would be zero. However, this is not the case in experiments, a signature that photons still keep some distinguishability even for zero delay. The parameter $\alpha$ takes into account this, reflecting the experimental visibility measured for coincidence counts as function of the temporal delay between photons. The parameter $\Delta=W_0/c$ is the width of the temporal fluctuations associated to the random changes of path imbalances.

Eq.~(\ref{sigma_qDIC}) tell us that the effect of random phase noise in quantum DIC is to reduce the visibility of the HOM effect as $\alpha \longrightarrow \alpha/\left(1+4 \sigma^2 \Delta^2\right)$. The bandwidth $\sigma$ in angular frequency and the bandwidth in wavelength $\Delta \lambda_s$ are related by $\sigma=\left( 2 \pi c/\lambda_s^2\right)\, \Delta \lambda_s$.
If we define the coherence length as $L_c=\lambda_s^2/\Delta \lambda_s$, we can write $1+4 \sigma^2 \Delta^2=1+16 W_0^2/L_c^2$. In quantum DIC the effect of random phase noise depends on the quotient $W_0/L_c$. As example, let us consider type II SPDC pumped with a quasi-monochromatic pump beam. The coherence length is $L_c= \Delta n_g L$, where $\Delta n_g$ is the difference of group refractive index for both polarizations and $L$ is the nonlinear crystal length. For the nonlinear crystal PPKTP at $810$ nm, $\Delta n_g \sim 0.1$. Therefore, we have that for $L=20$ mm, the coherence length is $L_c \sim 2$ mm. 

In Supplementary Section III we calculate the precision of phase estimation for classical DIC based on the setup shown in Figure~\ref{experimental_setup1}(a). We obtain that the precision is
\begin{equation}
\sigma_{\varphi}^2=\frac{1}{T_d N_0 M}\,\frac{T_1+T_2}{2 T_1 T_2}\, \frac{1}{\gamma^2 \cos^2 \varphi_0} + \frac{1}{2}\,\frac{1-\gamma^2}{M \gamma^2} \frac{1+\gamma^2 \cos \,2\varphi _0}{\cos^2 \varphi_0}
\label{sigma_cDIC}
\end{equation}
where $\varphi_0=k_0 (n_0-1)d$, and $\gamma=\exp \left[-2\pi^2 W_0^2/\lambda^2 \right]$. To obtain the precision in the estimation of the step height $d$, we make use of the relationship $\sigma_d/\sigma_{\varphi}=\lambda /\left[ 2 \pi (n_0-1) \right]$. Equation (\ref{sigma_cDIC}) tells us that the effect of random phase fluctuations on sensitivity depends on the value of $W_0/\lambda$, in contrast to the case of quantum DIC, where it depends on $W_0/L_c$. In most experiments, $L_c \gg \lambda_s$, which explains the enhanced resilience to random phase noise of quantum versus classical DIC.

\begin{figure*}[t!]
 \centering \includegraphics[clip,trim={2cm 2.7cm 0cm 1cm}, clip, width=\linewidth]{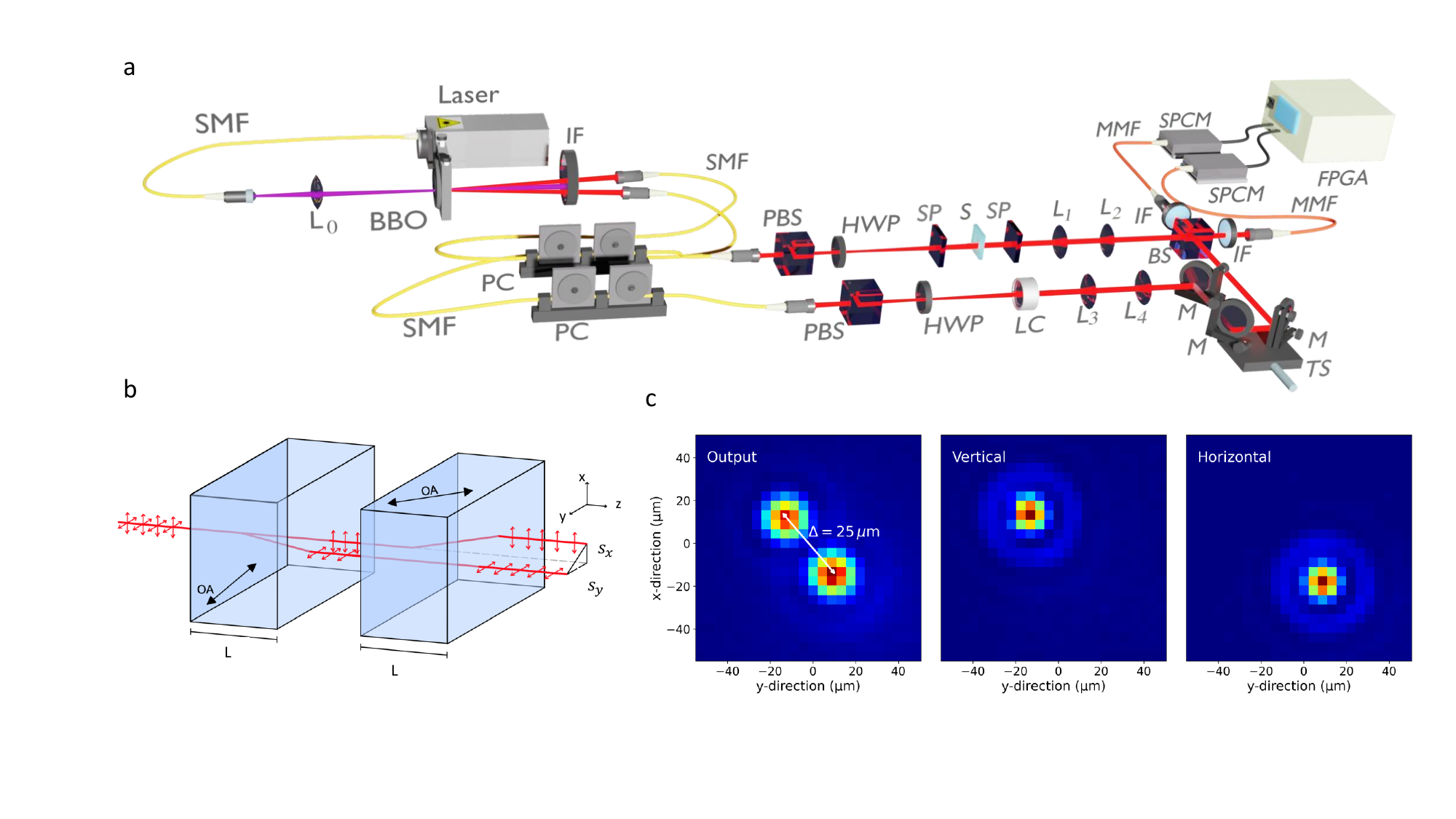}
 \caption{\textbf{ Experimental setup of the qDIC microscope and principle of quantum differential interference contrast (qDIC) microscopy using Savart plates.}
(a) Photon pairs generated via SPDC are prepared in diagonal polarization and sent to the signal and idler arms. The signal photon probes the sample placed between two Savart plates, while a liquid crystal in the idler arm introduces a tunable phase. Both photons interfere at a beam splitter, and coincidences are recorded after spectral filtering and fiber coupling. The sample is placed in a two-dimensional translation stage that permits the scan. (b) shows the working-principle of a Savart plate (SP): the first crystal displaces horizontally polarized light while leaving vertical polarization unchanged; the second crystal, rotated by 90°, displaces the vertical polarization, producing two spatially separated beams. (c) is the experimental images of a beam transmitted through a $10~\mu\text{m}$ pinhole and a $3~\text{mm}$ Savart plate, showing the relative displacement of the two polarization components.}
 \label{fig: experimentalsetup}
\end{figure*}

Figure \ref{qadvantage} shows a comparison between the precision of phase estimation attainable for both classical and quantum DIC as function of the amount of random phase noise present, for a representative case that we will consider in our experiments. In the ideal case with negligible random phase noise, classical DIC shows a better sensitivity than quantum DIC. However, even a modest amount of random phase noise, as small as $\sim \lambda/100$, reverse the situation and leads to quantum DIC showing enhanced sensitivity over classical DIC, even when a lower number of photons interacting with the sample in the quantum case. Typical values of the random phase noise present in {\it good} interferometers has been measured to be in the order of up to tens of miliradians \cite{noise2003,noise2024}.

\section*{Quantum differential interference contrast microscope setup}
Figure~\ref{fig: experimentalsetup}(a) shows the experimental implementation of the quantum Differential Interference Contrast microscope. A continuous-wave (CW) laser at 407~nm, coupled into a single-mode fiber (SMF), is used to pump a type I BBO nonlinear crystal with length $L=2~\text{mm}$. Lens $L_0$ focus the pump light onto the crystal. The nonlinear crystal is aligned to maximize the production of SPDC signal/idler photon pairs ($e \rightarrow oo$) with a half-opening angle of $3^{\circ}$. Signal/idler photons are collected into single-mode fiber (SMF) to remove all spatial correlations between the twin photon. 

In both the signal and idler arms, a fiber polarization controller (PC) and a polarizing beam splitter (PBS) placed after a collimator are used to set the polarization to horizontal. A half-wave plate (HWP) oriented at $22.5^{\circ}$ then rotates the polarization to diagonal. In the idler arm, after the HWP, a computer-controlled liquid crystal variable retarder (LCVR) device is used to control the polarization of the reference photons, introducing a phase shift between horizontal and vertical polarizations. A telescope (lenses $L_3$ and $L_4$) magnifies and collimates the idler beam. Two mirrors in this arm are mounted on a motorized translation stage to control the delay of idler photons. 

The cornerstone elements of the DIC technique are optical beam displacers such as the Savart plate (SP) and Wollaston prisms. In our experiments, the sample (S) is placed between two Savart plates in the signal photon arm, at the location of the beam waist. The second Savart plate is rotated by $90^{\circ}$ with respect to the first one. It is mounted on a two-dimensional translation stage for lateral scanning. A telescope (lenses $L_1$ and $L_2$) magnifies and collimates the signal beam after the second Savart plate. 

A Savart plate consists of two plates made of a birefringent material, such as quartz or calcite [see Fig.~\ref{fig: experimentalsetup} (b)]. The plates are arranged so that the first plate displaces horizontally polarized light, while vertically polarized light remains unaffected. The second plate is rotated relative to the first one so that the effect is reversed: vertically polarized light is displaced, while horizontally polarized light propagates without deviation. In this way, the horizontal polarization is displaced along the $x$-direction, and the vertical polarization is displaced along the $y$-direction, as shown in Fig.~\ref{fig: experimentalsetup}(c). 

Figure~\ref{fig: experimentalsetup}(c) shows an experimental example of the functioning of a Savart plate. The images correspond to light passing through a $10~\mu\text{m}$ pinhole and a Savart plate of length $L_{SP}=3~\text{mm}$. The two spots correspond to the displacements of the two orthogonal polarization components in different directions. The middle image shows the position of the spot with vertical polarization. A polarizer is placed in front of the camera to block the other spot with the orthogonal polarization. Rotating the polarizer by $90^{\circ}$ switches the transmitted polarization, revealing the spot corresponding to horizontal polarization, as shown in the right image. 

The signal and idler photons are directed to the two input ports of a beam splitter (BS). In each output port, an interferometric filter (IF) centered at 820~nm (tilted by $\sim 9^{\circ}$) is used to spectrally select the photons. The filtered photons are collected into multimode fibers (MMF).The detection is done using single photon detectors (SPD) and the coincidences and single counts are recorded using a field-programmable gate array (FPGA, Abacus, Tausand) with a coincidence window of $5~\text{ns}$.

\begin{figure}[t!]
 \centering
 \includegraphics[clip,trim={.8cm 0cm 17.5cm 0cm}, clip, width=\linewidth]{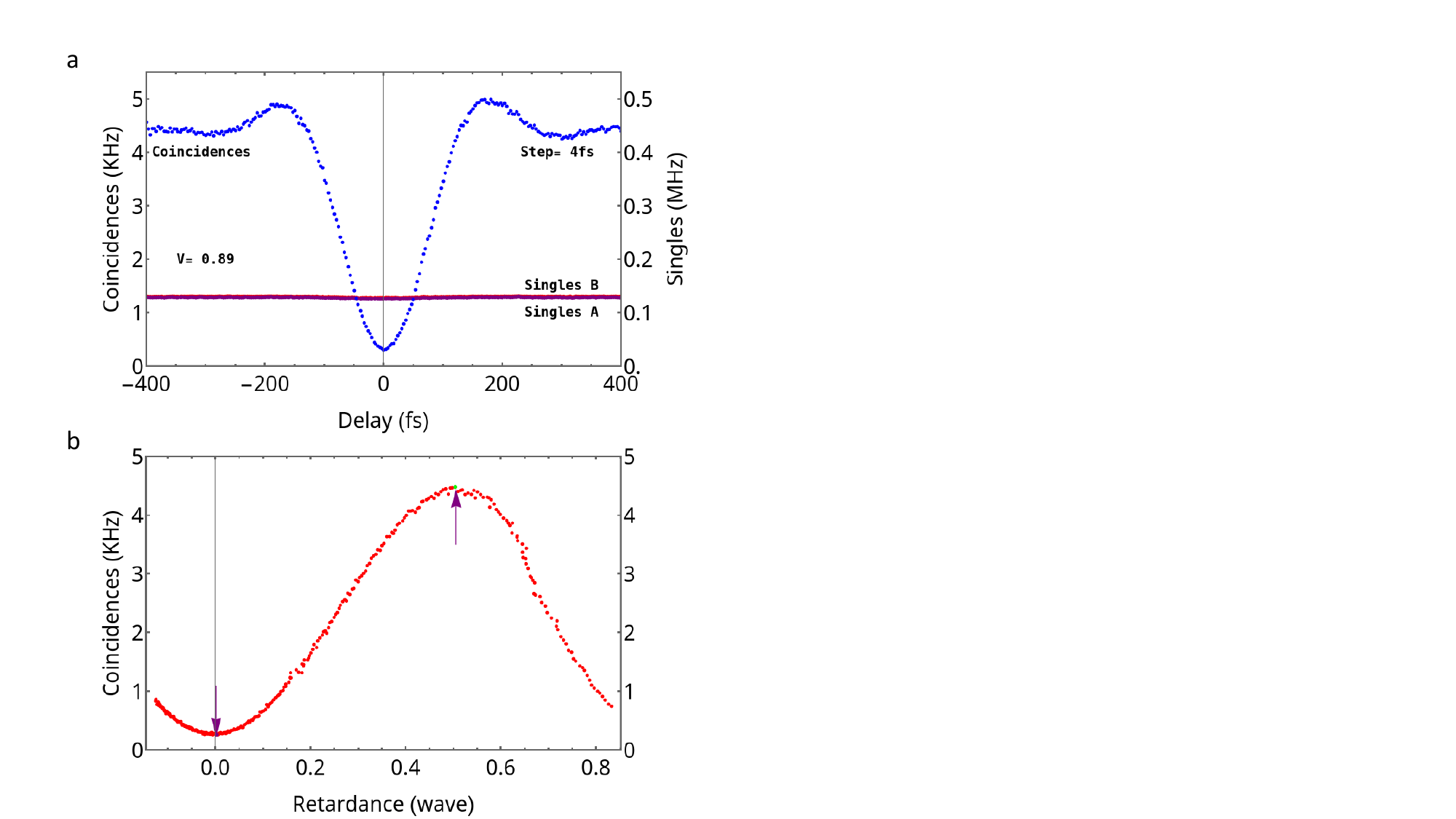}
 \caption{\textbf{Experimental HOM dip obtained by varying the temporal delay between photons and changing the polarization state of one photon}. (a) When the delay is zero, the photons are temporally indistinguishable and the coincidence rate $R_{cc}$ reaches its minimum, while single-photon counts remain constant. Increasing the delay makes the photons distinguishable, raising $R_{cc}$. (b) The coincidence rate $R_{cc}$ is controlled by a relative phase $\theta$, which determines the polarizations of the idler photon. When $\theta = 0$, polarizations of the photons are identical and coincidences counts almost vanish. When $\theta = \pi$, the polarizations are orthogonal and coincidences counts reach their maximum.}
\label{fig: dip}
\end{figure}

To demonstrate that we can compare how different are the polarizations of signal (sample arm) and idler photons (reference arm), we first need to make the temporal delays in both arms equal. For this, without the sample, the polarization of signal and idler photons is set to diagonal, and we measure the number of coincidence counts $R_{cc}$ measured as function of a variable temporal delay. Figure~\ref{fig: dip}(a) shows the experimental data obtained. When the temporal delay between sample and reference arms is nonzero, the photons are distinguishable and $R_{cc} \neq 0$. When the delay is zero, they are temporally indistinguishable and $R_{cc}$ reaches a minimum. The parameter $\alpha$ introduced in Eq.~(\ref{sigma_qDIC}) characterizes experimentally the residual distinguishability that still exist even for zero delay. Fig.~\ref{fig: dip}(a) also shows that the single-photon count rates at each output port remain constant, a signature that the dependence of coincidence counts versus delay is due to the degree of quantum interference that exists between signal and idler photons. 

Figure~\ref{fig: dip}(b) shows the number of coincidence counts as function of the polarization of the idler photons, modified changing the angle $\theta$. When signal and idler photons share the same state of polarization ($\theta=0$), there is a minimum of the coincidence counts. When they have orthogonal polarizations ($\theta=\pi$), there is a maximum.

\section*{Experimental results}
Before entering the beam splitter, if we assume zero delay between sample and reference arms and neglect the frequency dependence of photons, the spatio-temporal shape of the quantum state reads 
\begin{eqnarray}
 & & \ket{\Phi \left( {\bf r}_1,{\bf r}_2\right)} = \frac{1}{2} \int_S d{\bf r}_1 f_0 \left( {\bf r}_1 \right) \Big\{ a_H^{\dagger} \left( {\bf r}_1 \right)+ \exp \left[ i \theta \right] a_V^{\dagger} \left( {\bf r}_1\right) \Big\} \nonumber \\
 & & \times \int_S d{\bf r}_2 f_0 \left( {\bf r}_2 \right) \Big\{ a_H^{\dagger}\left( {\bf r}_2 \right)\, e^{i \varphi_H \left( {\bf r}_2 \right)} + a_V^{\dagger} \left( {\bf r}_2 \right)\,\, e^{i \varphi_V \left( {\bf r}_2 \right)} \Big\} 
\end{eqnarray}
where ${\bf r_1}=(x_1,y_1)$ ${\bf r_2}=(x_2,y_2)$ designate transverse coordinates in the two input ports of the beam splitter, $f({\bf r})$ is the input Gaussian-like spatial shape of signal and idler photons, $S$ is the detection area, $\varphi_{H,V}$ are the spatially dependent phase shifts acquired by signal photons after traversing the sample and $\theta$ is the controllable phase shift added to the vertical component of idler photons by the LCVR. When the signal and idler photons impinge simultaneously on the BS from different input ports, the probability to measure a coincidence count (see Supplementary Section IV for details) is
\begin{equation}
 P_{cc} = \frac{\gamma_0^2}{2} - \frac{|\alpha|^2}{8} \, |\gamma_{V} + \gamma_{H}|^2
 \label{probability100}
\end{equation}
where $\gamma_{0}=\int_S d{\bf r} |f_0({\bf r})|^2$, $\gamma_{H} = \int_S d\mathbf{\bf r} |f_0({\bf r})|^2\, \exp \left[ i \varphi_H \left( {\bf r}\right) \right]$ and $\gamma_{V} = \int_S d{\bf r} |f_0({\bf r})|^2\, \exp \left[ i \varphi_V \left( {\bf r}\right) \right]$ and $\alpha$ accounts for any residual distinguishability present in all other degrees of freedom. 

Variations in the topography of the sample thickness induce polarization changes in the signal photon, which in turn modify the coincidence count rate. This effect can be quantified by measuring the visibility, the comparison of the coincidence count rate using the idler polarization state that minimizes coincidences, and using the polarization that maximizes coincidences. The polarization of the idler photon is controlled via the LCVR located in its path, which controls the value of the phase $\theta$. If $P_{cc}^0$ designates the probability of observing a coincidence count for $\theta_0=0$, when the state of polarization of the idler photon before the beam splitter is diagonal, and $P_{cc}^{\pi}$ as the probability for $\theta=\pi$, when the state of polarization of the idler photon before the beam splitter is anti-diagonal, we can measure experimentally the visibility
\begin{equation}
\Gamma=\frac{P_{cc}^{\pi}-P_{cc}^{0}}{P_{cc}^{\pi}+P_{cc}^{0}}
 \label{visibility100}
\end{equation}
Making use of Eq. (\ref{probability100}), it is straightforward to show can that
\begin{equation}
 \Gamma=\frac{\gamma_H \gamma_V^{*}+\gamma_H^{*} \gamma_V}{4\gamma_0^2-|\gamma_H|^2-|\gamma_V|^2}
\end{equation}
In the absence of an object in the sample arm, $\gamma_0 = \gamma_H = \gamma_V$ and the visibility is $\Gamma = 1$. For a sufficiently large detection area such that all photons are collected, $\gamma_0 = 1$. If the area $S$ around the position ${\bf r}_0$ is small enough so that the phases $\varphi_H ({\bf r})$ and $\varphi_V ({\bf r})$ can be assumed to be constant inside the detection area $S$, we have $\varphi_0 ({\bf r}_0)=\cos^{-1}\, \Gamma ({\bf r}_0)$. This is the tool we use in experiments to determine the value of the unknown phase.

\begin{figure}[t!]
 \centering

 \includegraphics[ clip,trim={0cm 0cm 0cm 0cm}, clip, width=\linewidth]{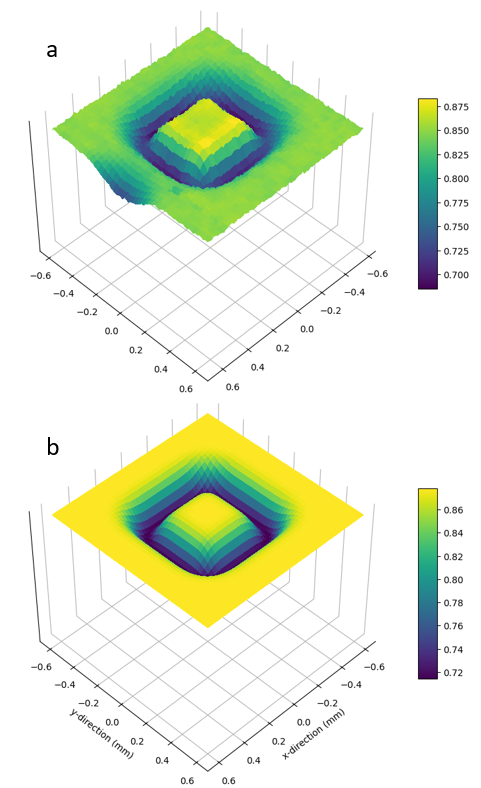}
 \caption{\textbf{Experimental results obtained with the quantum DIC microscope} Upper figure is the 2D topography measured for the sample with step height $h=44~\text{nm}$. The bottom figure shows the 2D theoretical prediction}
 \label{fig: Results}
\end{figure}

\begin{figure}[t!]
 \centering
 \includegraphics[angle=-90, clip,trim={0cm 0cm 7.5cm 0cm}, clip, width=\linewidth]{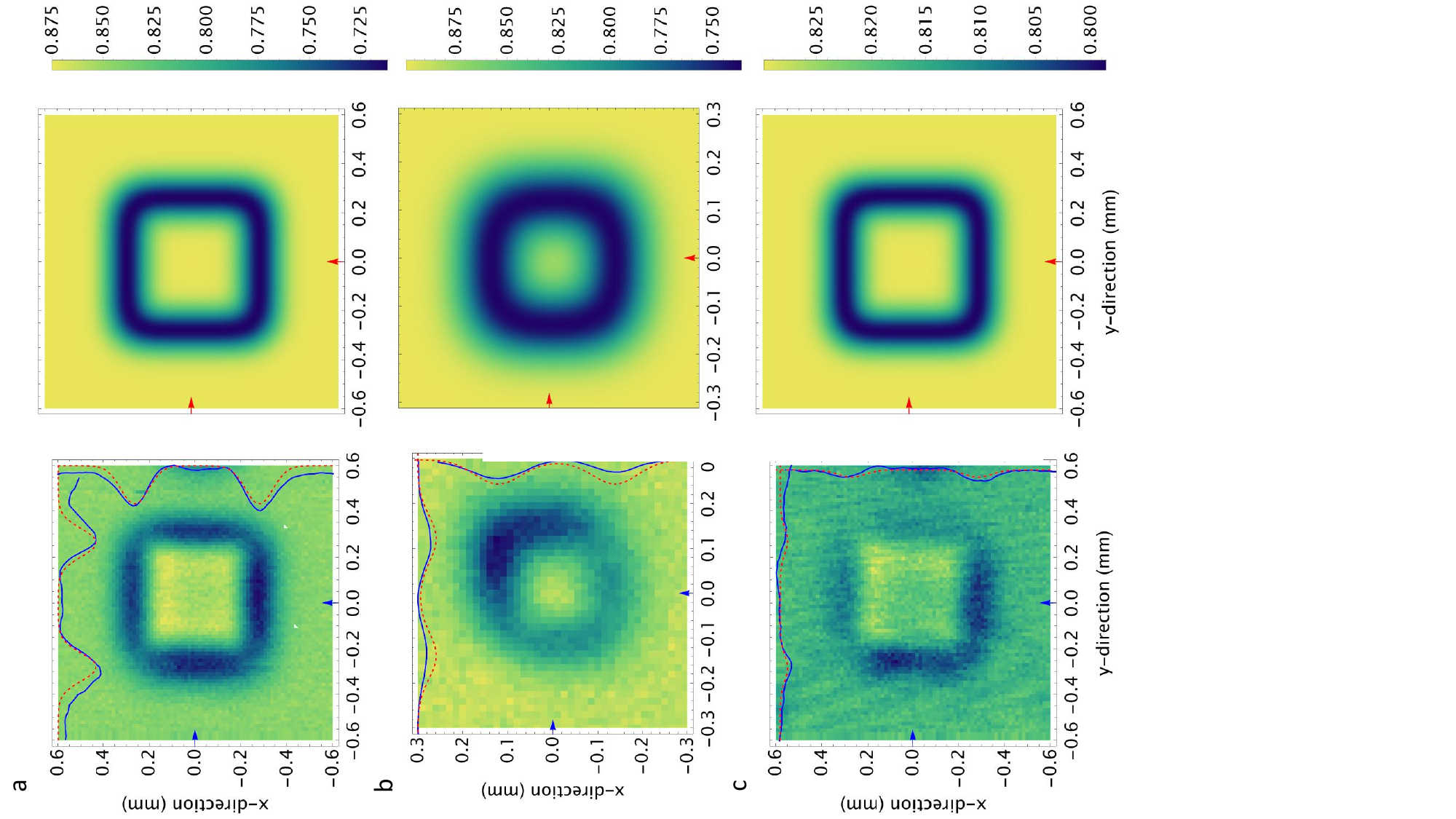}
 \caption{\textbf{Experimental results obtained with the quantum DIC microscope.} The images in the left column are the experiment results, and the images in the right column are the corresponding theoretical predictions considering the nominal values of the samples. The first row shows images for a sample with step height of $h=44~\text{nm}$. The second row is for $h=77~\text{nm}$ and last one is $h=173~\text{nm}$. The lines on the sides of each experimental plot correspond to a 1D cut in the center of the sample.}
 \label{fig: 3dResults}
\end{figure}

Making use of Eq.~(\ref{visibility100}), we perform measurements of the topography of the three samples, shown in Figs.~\ref{fig: Results} and \ref{fig: 3dResults}. The samples are cliff-like structures made in $\text{SiO}_2$ with different step heights, with nominal values of $h=44$, $77$ and $173$ nm. The shape of the optical trench is a square with lateral size $s=500 \mu$m (for samples with $h=44$ and $173$ nm) and $s=250 \mu$m (for the sample with $h=77$ nm). 

In Fig.~\ref{fig: 3dResults}, the left image corresponds to a 2D experimental scan and the right image is the corresponding theoretical prediction. The simulations are done by employing average of the values of the height obtained in the characterization of the sample using AFM microscopy. The colormap in both figure corresponds to the value of the visibility $\Gamma$. The profiles (red dash line: theory and blue solid line: experiments) correspond to the positions that the arrows point out. The discrepancy between the experimental scan and the simulation is due to non-uniform thickness across the structure and to the edge irregularities due to fabrication imperfections that are unveiled by the Atomic Force Microscopes images taken (see Methods and Supplementary Section V). 

\section*{Discussion}
We have demonstrated a quantum version of a Differential Interference Contrast (qDIC) microscope. Similarly to a classical DIC microscope, optical phase gradients of a sample translates into changes in the polarization state of photons, which need to be measured. However, polarization analysis in quantum DIC makes use of a inherently quantum phenomenon, quantum coherence in a Hong-Ou-Mandel scheme. While the HOM effect is insensitive to phase differences, the DIC scheme maps spatially dependent phase variations into distinct polarization states, rendering them accessible to HOM interference. This synergy between DIC and HOM thus provides a direct quantum pathway for phase-sensitive imaging.

We have shown that quantum DIC is more resilient to the presence of random phase noise in the interferometer than classical DIC. In our experiments, we can measure nanometric features of samples without making use of any optical or electronic systems to stabilize the phase of the interferometer. This reveals the robustness of quantum DIC under the presence of random phase noise. There are other systems to mitigate the harmful effects of random phase noise, but they are generally implemented at the expense of easiness and simplicity of the optical systems, sought-after qualities {\it much} appreciated for practical applications.

In order to keep the beneficial effects of quantum concepts in all types of applications, quantum technologies generally need to set up control systems to mitigate the harmful effects coming from all kinds of noise disturbances and losses. There is a striking contrast between this fact and translating classical DIC to the quantum world. In this case, quantum DIC is more resilient to the presence of random noise effects in an interferometer than classical DIC, when one would initially expect the contrary. 

Finally, another key advantage of our approach is that no post-processing of the HOM signal is required to reconstruct the phase gradient of the sample. The information is directly encoded in the coincidence rate, offering a resource-efficient and intrinsically quantum method for phase retrieval. This establishes qDIC microscopy as a promising tool for non-invasive, high-sensitivity imaging of transparent and weakly scattering samples.

\section*{Method}
{\small \subsection*{Protocol}}
With the temporal delay between photons and the polarization appropriately controlled, we can image a sample with the quantum Differential Interference Contrast microscope implemented. A sample is placed between the two Savart plates. We check again what is the delay time required in the idler path, since the presence of the sample could introduce extra temporal delays. The experimental scans measure the number of coincidences between photons in different output ports of the BS, while the sample is displaced. The sample is moved in steps $100~\mu \text{m}$ in the transverse plane with respect to the direction of propagation of the signal photons. 

Coincidences were measured when the retardance introduced by the LCVR is $0$ and $0.5\lambda$, which yields the maximum coincidence rate $R_{max}$ and the minimum value of $R_{min}$ for each transverse position of the sample. The coincidence rate was measured by counting the events in a coincidence window of $5~\text{ns}$ during $1$ second, repeating the measurement ten times and averaging the results. Using these numbers we calculate the value of the parameter $\Gamma$. The whole transverse scan allows the retrieval of the phase gradient of the sample.

{\small \subsection*{Characterization via AFM}
Atomic force microscopy (AFM) topographies were obtained with a NX20 microscope (Park Systems) using the Non-Contact mode to avoid possible morphological/mechanical damage in the samples. PPP-NCHR-20 (from Nanosensors) tips were used for the scan with a resonance frequency of 305~kHz, 12~nm of tip amplitude and scan area of $50\times50~\mu$m. All the images were obtained in the same conditions. Image processing and data analysis were performed with the Gwyddion open source software (v. 2.56), where mean plane subtraction, Polynomial filter and profiles were determined. Supplementary Section V shows a typical AFM photograph of one of the samples, which highlights the difference between the nominal and the real topography we observe and measure.}

\bibliography{References}

\section*{Acknowledgments}
We acknowledge support from the project NOVISLIGHT (PID2023-149780NB-I00) funded by Ministerio de Ciencia, Innovaci\'on y Universidades (Proyectos de generaci\'on de conocimiento 2023). 
This work is part of the R$\&$D project CEX2024-001490-S, funded by MCIN/ AEI/10.13039/501100011033/. We acknowledge support from Fundació Cellex, Fundació Mir-Puig, and from Generalitat de Catalunya through the CERCA program. DFU acknowledge financial support (PRE2018-085072) financed by MCIN/AEI/10.13039/501100011033 and FSE ``El FSE invierte en tu futuro''

\section*{Author contributions}
JPT and DFU came with the idea of a quantum Differential Interference Contrast (DIC) microscope and designed the experiment to demonstrate it. They also made the theoretical calculations to understand and explain how it works. DFU and MHMP performed the experiments. FBT characterized the samples used in our experiments. JPT and DFU wrote the manuscript, with the help of MHMP.

\section*{Additional information}

\section*{Competing financial interests}

The authors declare no competing financial interests.
\end{document}